\documentclass[12pt]{iopart}
\usepackage{graphicx}
\usepackage{iopams}

\newcommand{\ket}[1]{|#1\rangle}
\newcommand \be{\begin{equation}}
\newcommand \ee{\end{equation}}
\newcommand \bea{\begin{eqnarray}}
\newcommand \eea{\end{eqnarray}}
\newcommand \bse{\begin{subequations}}
\newcommand \ese{\end{subequations}}

\begin{document}

\title[Coherent control of mesoscopic atomic ensembles]{Coherent control of mesoscopic atomic ensembles for quantum information}

\author{I~I~Beterov$^{1,2}$, M~Saffman$^3$, V~P~Zhukov$^4$, D~B~Tretyakov$^1$, V~M~Entin$^1$, E~A~Yakshina$^{1,2,5}$, I~I~Ryabtsev$^{1,2,5}$, C~W~Mansell$^6$, C~MacCormick$^6$, S~Bergamini$^6$, M~P~Fedoruk$^{2,4}$}

\address{$^1$A.V.Rzhanov Institute of Semiconductor Physics SB RAS, 630090 Novosibirsk, Russia }
\address{$^2$Novosibirsk State University, 630090 Novosibirsk, Russia}
\address{$^3$Department of Physics, University of Wisconsin, Madison, Wisconsin, 53706, USA}
\address{$^4$Institute of Computational Technologies SB RAS, 630090 Novosibirsk, Russia}
\address{$^5$Russian Quantum Center, Skolkovo, Moscow Reg., 143025, Russia}
\address{$^6$The Open University, Walton Hall, MK7 6AA, Milton Keynes, UK}

\ead{beterov@isp.nsc.ru}

\begin{abstract}
We discuss methods for coherently controlling mesoscopic atomic ensembles where the number of atoms varies randomly from one experimental run to the next. The proposed schemes are based on adiabatic passage and Rydberg blockade and can be used for implementation of a scalable quantum register formed by an array of randomly loaded optical dipole traps. 
\end{abstract}

\maketitle

\section{Introduction}

\begin{figure}[!t]
\includegraphics[width=\columnwidth]{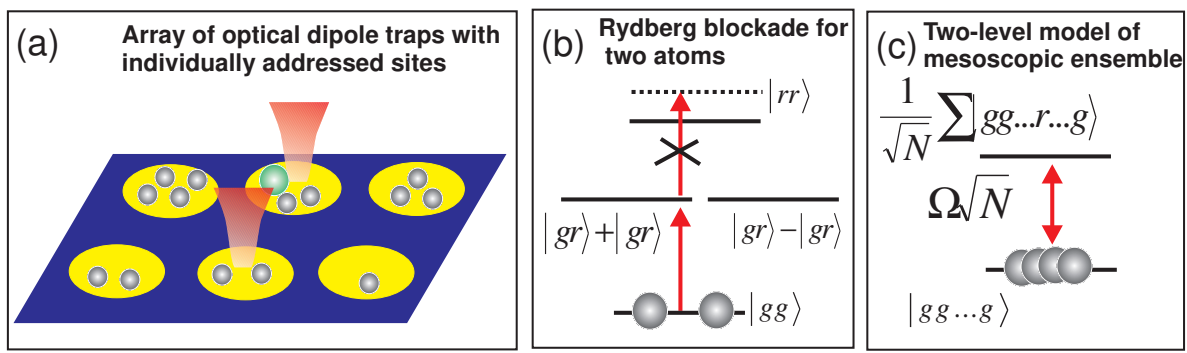}
\vspace{-.5cm}
\caption{
\label{RegisterScheme}(Color online).
(a) Scheme of the quantum register based on individually addressed atomic ensembles in the array of optical dipole traps. Laser pulses are used to excite atoms into the Rydberg state. Only one atom in each site can be excited due to Rydberg blockade. Simultaneous excitation of Rydberg atoms in the neighboring sites is also blocked; (b) Collective states of two interacting atoms. The shift of the collective energy level when both atoms are excited into the Rydberg state leads to suppression of double Rydberg excitation, known Rydberg blockade; (c) The mesoscopic ensemble of $N$ atoms in the Rydberg blockade regime can be considered as a two-level system with enchanced coupling to the laser field.
}
\end{figure}

Arrays of optical dipole traps are extremely promising for implementation of the scalable quantum register with neutral atoms \cite{Piotrowicz2013}. Deterministic single-atom loading of optical dipole traps is required for schemes of quantum computing where each atom is considered as a single qubit. Another approach to build a quantum register is based on the encoding of quantum information in collective states of ensembles of strongly interacting atoms in the regime of Rydberg blockade~\cite{Lukin2001}, as shown in figure~\ref{RegisterScheme}(a). Rydberg blockade manifests itself as a suppression of the excitation of more than one atom in the ensemble by narrow-band laser radiation due to the shifts of the collective  energy levels induced by long-range Rydberg-Rydberg interactions~\cite{Lukin2001}, as illustrated in figure~\ref{RegisterScheme}(b) for two atoms. The mesoscopic atomic ensemble in the Rydberg blockade regime can be considered as a two-level system with the coupling to the laser radiation enchanced by a factor of  of $\sqrt N$, with $N$ the number of atoms, as shown in figure~\ref{RegisterScheme}(c). This behavior has been experimentally demonstrated in experiments with several hundreds of atoms~\cite{Dudin2012}. Such enchancement is advantageous, since it allows implementation of fast quantum gates at moderate laser intensities, but it also leads to sensitivity of the quantum gate fidelity to fluctuations of the numbers of atoms. Although there is recent progress in nondestructive  measurement of  $N$ with high accuracy~\cite{HZhang2012} and control of the number of atoms in the ensemble~\cite{Ates2013}, it remains an outstanding challenge to implement high fidelity quantum logic gates without precise knowledge of $N$.

We have developed a method of deterministic single-atom Rydberg excitation based on adiabatic passage and Rydberg blockade~\cite{Beterov2011}. Combined with the proposal of \cite{Saffman2002} it allows for deterministic single-atom loading of the optical dipole traps and optical lattices. Due to high efficiency of Rydberg blockade our method has potential of overcoming the accuracy of single-atom loading based on laser-assisted collisions where the accuracy of 91\% has been experimentally demonstrated \cite{Carpentier2012}. 
However, implementation of a quantum register based on mesoscopic ensembles could be advantageous compared to single atoms due to reduced sensitivity to losses of atoms in the traps. We have developed the schemes of single-qubit and two-qubit quantum logic operations with mesoscopic ensembles containing random number of atoms based on double sequences of adiabatic Rydberg excitation with compensation of the dynamic phase~\cite{Beterov2013}.

Recently it has been proposed to use mesoscopic atomic ensembles on an atom chip for creation of cluster states which are required for measurement-based quantum computation~(MBQC)~\cite{Bin-Bin2010}. The MBQC scheme, first discussed in reference~\cite{Raussendorf2001}, is a sequence of destructive measurements performed on a register, which had been initially prepared in a so-called cluster state by Hadamard rotations of each qubit followed by the controlled phase operations between all nearest neighbours~\cite{Raussendorf2001}. The quantum algorithms are defined by the geometry of the register and order of measurements. Neutral atoms are perfectly suitable for MBQC due to the availability of fast and reliable method of destructive measurement of the quantum state of the atom via excitation to the Rydberg state with subsequent selective field ionization~\cite{Ducas1975,Gallagher1977}. We discuss the schemes of single-qubit rotations and controlled phase gate based on Rydberg interaction between the atomic ensembles located in the neighboring dipole traps. Combination of two-qubit gates between nearest neighbours, arbitrary single-qubit rotations and fast destructive measurement must be enough for implementation of the scalable MBQC with mesoscopic ensembles of neutral atoms. 

This article is organized as follows. Section 2 is devoted to single-photon and two-photon deterministic single-atom Rydberg excitation based on adiabatic passage and Rydberg blockade. Section 3 presents the protocols for single-qubit rotations, CNOT and controlled phase gates. 

\section{Deterministic single-atom excitation}

\begin{figure}[!t]
\includegraphics[width=\columnwidth]{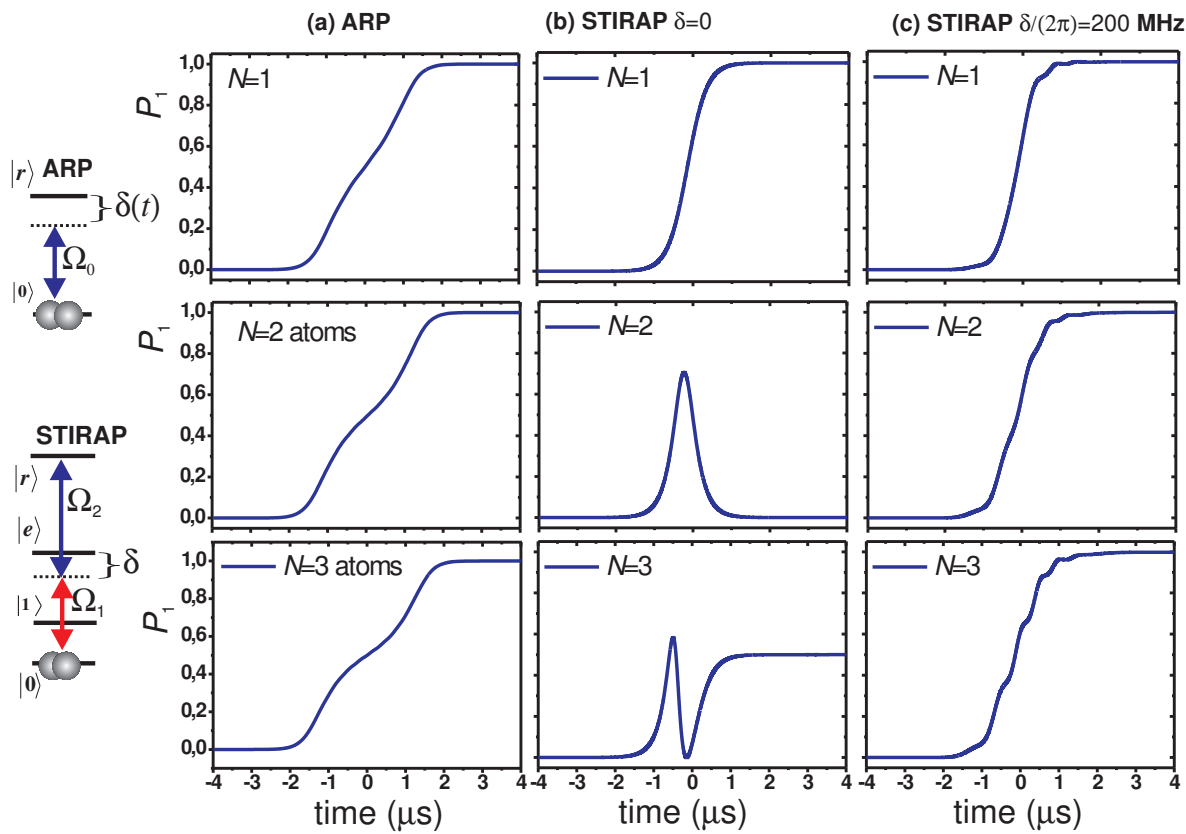}
\vspace{-.5cm}
\caption{
\label{ExcitationProbability}(Color online).
Calculated time dependence of the probability of single-atom Rydberg excitation for $N=1-3$ atoms (top to bottom).
(a) ARP with the chirp rate is  $\alpha/\left(2\pi\right)=1$~THz/s and Rabi frequency is $\Omega_1/\left(2 \pi \right)=2$~MHz;
(b) STIRAP with $\Omega_1/2\pi = 30~\rm MHz,$ $\Omega_2/2\pi = 40~\rm MHz$, $\delta/2\pi = 0$;
(c) STIRAP with $\Omega_1/2\pi = 30~\rm MHz,$ $\Omega_2/2\pi = 40~\rm MHz$, $\delta/2\pi = 200$~MHz.
}
\end{figure}

We have proposed to use adiabatic rapid passage (ARP) by chirped laser pulses or Stimulated Rapid Adiabatic Passage (STIRAP) to deterministically excite a single Rydberg atom in the regime of Rydberg blockade. The energy level scheme for single-photon ARP and two-photon STIRAP is shown in  figure~\ref{ExcitationProbability} (left).

We have numerically calculated the probability of single-atom Rydberg excitation in the mesoscopic ensembles with $N<10$ atoms in the regime of Rydberg blockade for a linearly-chirped 
Gaussian laser pulse and STIRAP sequence.  Calculations were performed using the Schr\"odinger equation, neglecting spontaneous emission, and assuming
perfect blockade so only states with at most a single Rydberg excitation were included. This model provides good agreement with our previous simulations of resonant dipole-dipole interaction and Rydberg blockade~\cite{Ryabtsev2010, Ryabtsev2010a}. In the time domain the electric field of the chirped pulse is expressed
as
\be
\label{eq1}
E\left({t} \right) = E_{0} \mathrm{exp}\left[ {\frac{{ - t^{2}}}{{2\tau ^{2}}}} 
\right]\mathrm{cos}\left[ {\omega _{0} t + \alpha \frac{{t^{2}}}{{2}}} \right].
\ee
Here $E_0$ is the peak electric field at $t = 0$, $\omega_{0}$ is the frequency of the atomic transition, $\tau = 1\,\mu \mathrm{s}$ is 
the half-width at $1/e$ intensity, and $\alpha $ is the 
chirp rate~\cite{Malinovsky2001}. We choose $E_{0} $ to be such as to provide a single-atom 
peak Rabi frequency $\Omega_1 /\left( {2\pi}  \right) = 2$~MHz or $\Omega_1 
/\left( {2\pi}  \right) = 0.5$~MHz. For convenience, the central frequency of the laser 
pulse is taken to be exactly resonant with the atomic transition at the 
maximum of the pulse amplitude. The atoms begin to interact with the laser 
radiation at $t = - 4\;\mu \mathrm{s}$.
The STIRAP sequence used 
\be
\label{eq2}
\Omega_{j}(t)=\Omega_{j} e^{-(t+t_j)^2/2\tau^2.}
\ee
for $j=1,2$ with $\Omega_1/2\pi = 30~\rm MHz,$
$\Omega_2/2\pi = 40~\rm MHz,$
$t_1=3.5~\mu\rm s$, $t_2=5.5~\mu\rm s$, $\tau=1~\mu\rm s,$ and $\delta/2\pi=200~\rm MHz$ or $\delta/2\pi=0$.

The numerically calculated time dependencies of the probability $P_1$ to excite a single Rydberg atom by the chirped laser pulse in the ensemble of $N$=1-3 atoms are shown in figure~\ref{ExcitationProbability}(a) for  ARP, in  figure~\ref{ExcitationProbability}(b)   for STIRAP with $\delta/(2\pi)=0$ and in figure~\ref{ExcitationProbability}(c) for $\delta/2\pi=200$~MHz. For ARP and STIRAP with large detuning $\delta/(2\pi)=200$~MHz from the intermediate state the probability of single-atom excitation is independent of the number of atoms, while for STIRAP with zero detuning from the intermediate state the regime of deterministic excitation breaks down for $N >1$~\cite{Beterov2011, Moller2008}. 

\begin{figure}[!t]
\includegraphics[width=\columnwidth]{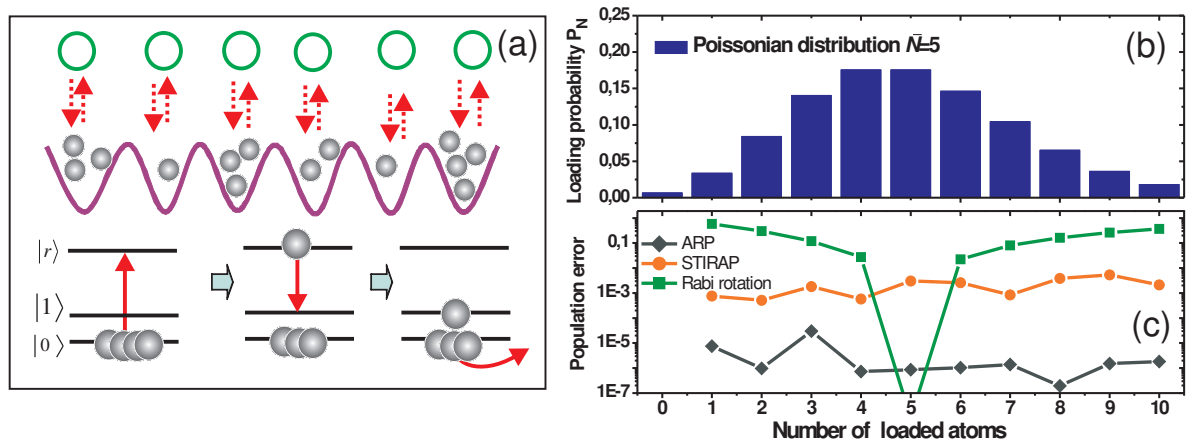}
\vspace{-.5cm}
\caption{
\label{Loading}(Color online).
(a) The scheme of single-atom loading using deterministic single-atom Rydberg excitation;
(b) The Poissonian statistics of loading of optical dipole trap witn the average number of atoms $\bar N=5$;
(c) The fidelity of single-atom excitation for a single-photon $\pi$ rotation with the area optimized for $N=5$ atoms compared
to STIRAP or ARP pulses.
}
\end{figure}

This technique of single-atom excitation can be used for deterministic single-atom loading, proposed in~\cite{Saffman2002}, when one of the atoms is deterministically transfered from between the hyperfine sublevels of the ground state through temporarily Rydberg excitation in the blockade regime, while all atoms remained at the inititally populated hyperfine sublevel are removed from the optical dipole trap by an additional laser pulse, as shown in figure~\ref{Loading}(a). A similar problem has been recently addressed in~\cite{Petrosyan2013}.
The probability of loading $N$ noninteracting atoms in a small  optical or magnetic trap is described, in general, by
Poissonian statistics. For $\bar{N}=5$ the probability to load zero atoms is 0.0067, as shown in figure~\ref{Loading}(b), which is  small enough to create a
large quantum register with a small number of defects. Figure~\ref{Loading}(c) shows a comparison of the
fidelity of single-atom excitation for a single-photon $\pi$ rotation with the area optimized for $N=5$ atoms compared
to STIRAP or ARP pulses. We see that the adiabatic pulses reduce the population error by up to several orders of
magnitude for a wide range of $N$.

\begin{figure}[!t]
\includegraphics[width=\columnwidth]{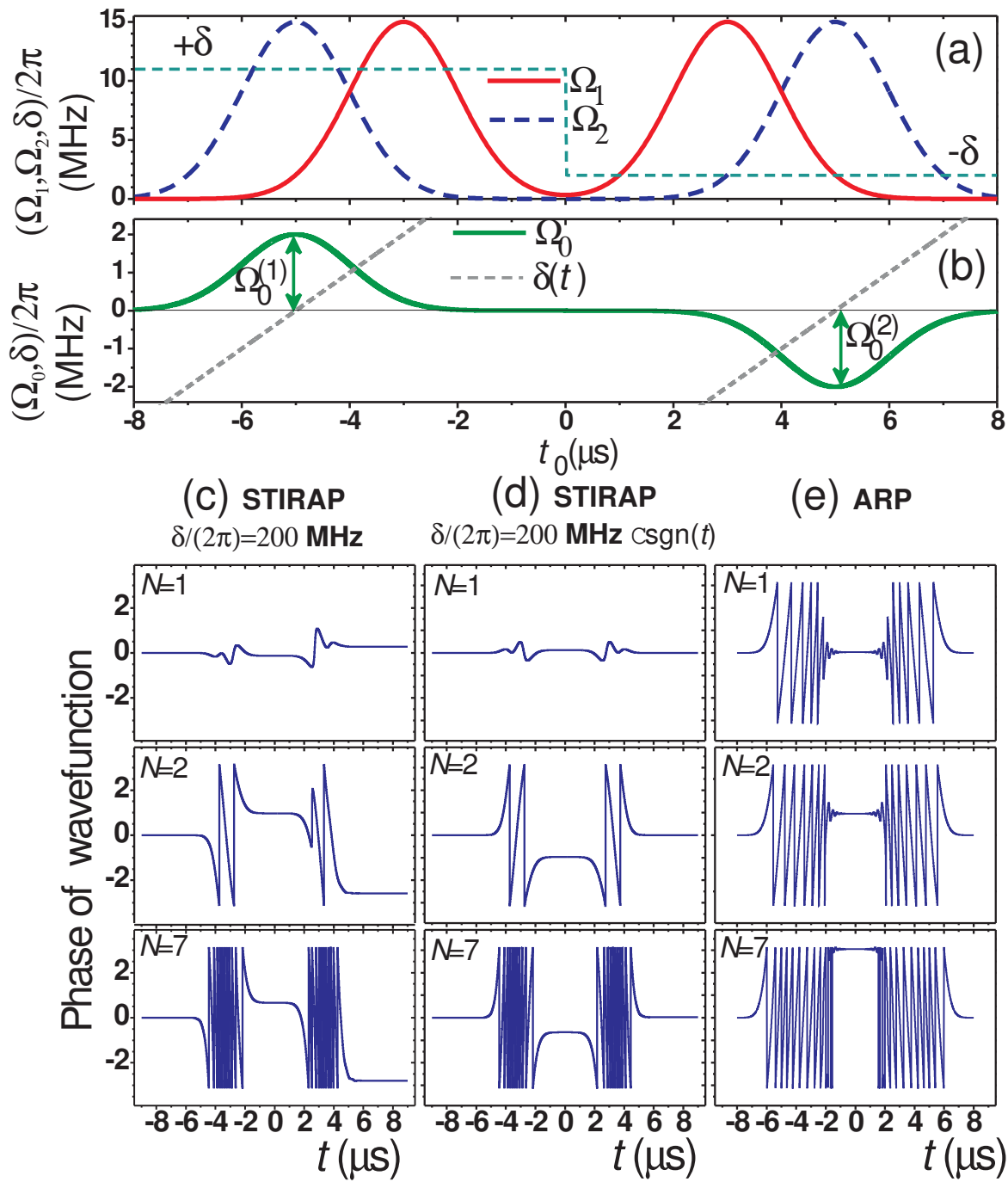}
\vspace{-.5cm}
\caption{
\label{DoubleSequence}(Color online).
(a) Time sequence of
STIRAP laser pulses; (b) Time sequence for ARP laser excitation; (c), (d), (e) The calculated time dependence of the phase 
of the collective ground state amplitude for $N=1,2,7$ atoms (top to bottom) for double STIRAP sequence (c)  with  $\delta/2\pi = 200~\mathrm{MHz}$, (d) with  $\delta/2\pi = 200~\mathrm{MHz} \times \mathrm{sgn}\left( t \right)$, and (e) for a double ARP pulse sequence with phase inversion.
}
\end{figure}

The accumulation of $N$-dependent dynamic phase during the adiabatic passage is the major obstacle for for implementation of quantum logic based on deterministic single-atom Rydberg excitation. Another difficulty is the inability to coherently transform the initially prepared superpositions of quantum states using adiabatic passage. Both problems have been addressed in the schemes of quantum gates which we have developed. We have proposed to use double ARP and STIRAP sequences for compensation of the dynamic phase, as shown in figure~\ref{DoubleSequence}(a) and (b). We have found that the phase of the atomic wavefunction can be compensated by switching the sign of the detuning between two STIRAP
pulses, or by switching the phase between two ARP pulses, as shown in figure~\ref{DoubleSequence}(a). For a double STIRAP sequence
with the same detuning throughout the accumulated phase depends on $N$ [figure~\ref{DoubleSequence}(c)], while the phase change
is zero, independent of $N$, when we switch the sign of detuning $\delta$ between the two STIRAP sequences
[figure~\ref{DoubleSequence}(d)]. A similar phase cancellation  occurs for $\pi$ phase shifted  ARP pulses
[figure~\ref{DoubleSequence}(e)], which can be implemented using an acousto-optic modulator.

\begin{figure}[!t]
\includegraphics[width=\columnwidth]{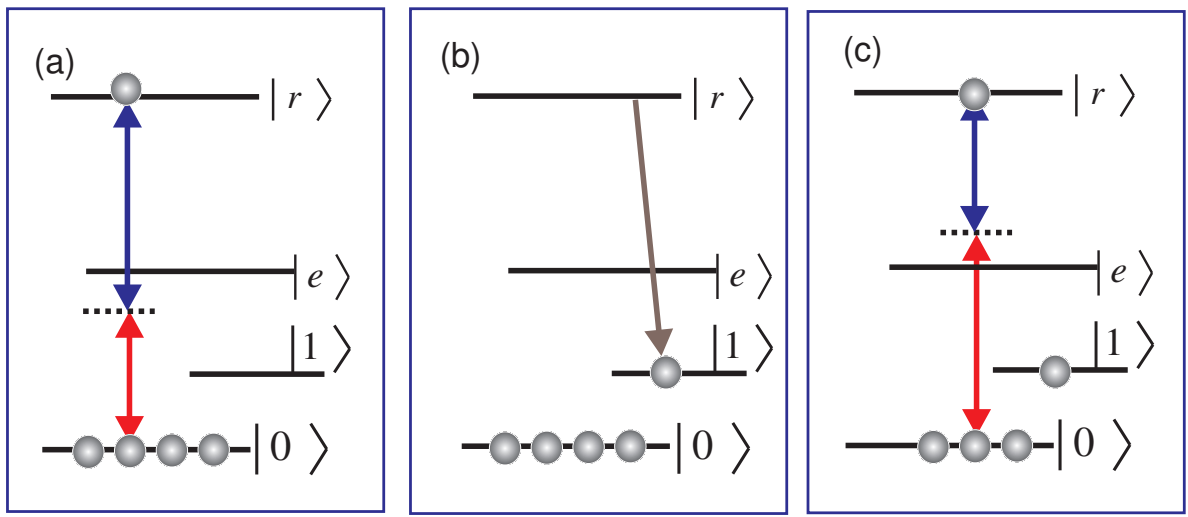}
\vspace{-.5cm}
\caption{
\label{Encoding}(Color online).
The sequence of STIRAP pulses with added $\pi$ rotation of the qubit. (a) Deterministic single-atom Rydberg excitation; (b) Transfer of the Rydberg excitation into the collective state with a single excitation of state $\ket{1}$ shared between all atoms of the ensemble; (c) Undesirable Rydberg excitation during the reverse STIRAP sequence;
}
\end{figure}

\section{Quantum gates based on adiabatic passage in mesoscopic ensembles}

We have developed protocols to implement quantum logic gates using phase compensated double STIRAP or
ARP.  Consider atoms with levels $\ket{0},\ket{1},\ket{e},\ket{r}$ as shown in figure~\ref{Encoding}. A qubit can be encoded in an $N$
atom ensemble with the logical states~\cite{Lukin2001} 
\bea
\ket{\bar 0} = \ket{000...000}, \\\nonumber
\ket{\bar 1}' =\frac{1}{\sqrt{N}}\sum_{j=1}^N \ket{000 ... 1_j ...000}. 
\eea

\noindent Levels $\ket{0},\ket{1}$ are atomic hyperfine ground states.
Coupling between these states is mediated by the singly excited Rydberg state 
\be
\ket{\bar r}' =\frac{1}{\sqrt{N}}\sum_{j=1}^N \ket{000 ... r_j ...000}.
\ee
\noindent
Rydberg blockade only allows single excitation of  $\ket{r}$
so the states $\ket{\bar 0}$ and $\ket{\bar r }'$ experience a collectively enhanced coupling rate $\Omega_N=\sqrt
N\Omega$. States $\ket{\bar r}'$ and $\ket{\bar 1 }'$ are  coupled at the single atom rate $\Omega$. State $\ket{\bar
1}'$ is produced by the sequential application of $\pi$ pulses $\ket{\bar 0}\rightarrow\ket{\bar r}'$ and $\ket{\bar
r}'\rightarrow\ket{\bar 1}'$, as shown in figure~\ref{Encoding}(a) and (b). Since the collective state $\ket{\bar 1}$ in the ensemble with $N>1$ atoms has $N-1$ 
atoms in state $\ket{0}$, the second STIRAP or ARP sequence used for phase compensation will lead to an undesirable single-atom Rydberg excitation at the end of the gate operation, as shown in figure~\ref{Encoding}(c). This collective state is expressed as 
\be
\ket{\bar r_a}' =\frac{1}{\sqrt{N(N-1)}}\sum_{j=1}^N\sum_{k\neq j} \ket{000 ... 1_j ...r_k...000}. 
\ee

\begin{figure}[!t]
\includegraphics[width=\columnwidth]{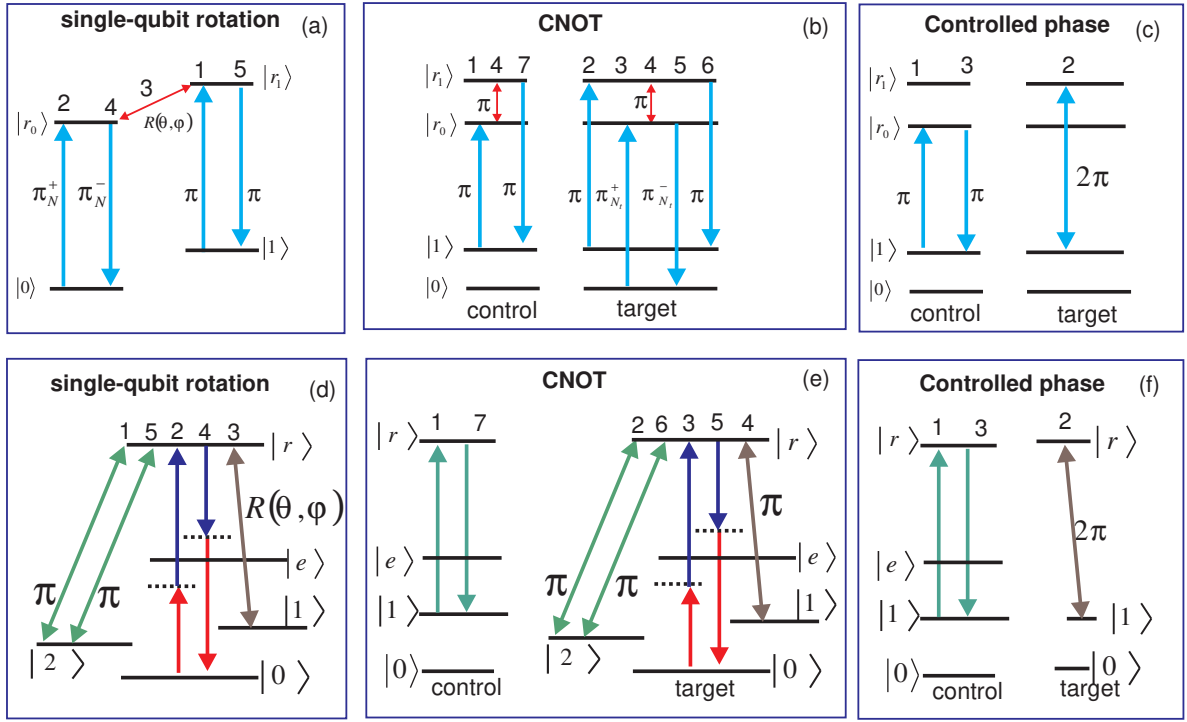}
\vspace{-.5cm}
\caption{
\label{Gates}(Color online).
The quantum gates for a mesoscopic  qubit with $N$ atoms. Microwave gates: (a) Single-qubit rotation; (b) CNOT; (c) Controlled phase gate. All-optical gates: (d) Single-qubit rotation; (e) CNOT; (f) Controlled phase gate. 
}
\end{figure}

\noindent
In order to solve this problem, we have developed  two general schemes which are based on two different structures of atomic energy levels. In the first scheme, shown in figure~\ref{Gates}(a)-(c) we use two hyperfine sublevels of the ground state of alkali-metal atom $\ket{0},\ket{1}$ for storage of quantum information and two $\ket{r_0},\ket{r_1}$ auxiliary Rydberg states coupled by the microwave radiation for coherent rotation of the ensemble qubit on arbitrary angles after Rydberg excitation. In the second scheme, shown in figure~\ref{Gates}(d)-(f) we use three magnetic subleveles of the hyperfine state, marked as  $\ket{0}, \ket{1}, \ket{2}$ for storage of quantum information. The sublevel  $\ket{2}$ is an auxiliary state, required to remove undesirable Rydberg excitation of ground-state atoms after the end of the gate operation. 
We consider these schemes separately. 

\textit{Microwave gates}

We define the ensemble states as:
\bea
\ket{\bar 0} &=&\ket{000...000}\\\nonumber
\ket{\bar 1}' &=& \frac{1}{\sqrt{N}}\sum_{j=1}^N \ket{000 ... 1_j ...000}\\\nonumber
\ket{\bar r_0}' &=& \frac{1}{\sqrt{N}}\sum_{j=1}^N \ket{000 ... (r_0)_j ...000}\\\nonumber
\ket{\bar r_1}' &=& \frac{1}{\sqrt{N}}\sum_{j=1}^N \ket{000 ... (r_1)_j ...000}.
\eea

\noindent The basic idea of our gate, shown in figure~\ref{Gates}(a) is to block undesirable Rydberg excitation by transfer of the population of the initially excited Rydberg state $\ket{r_0}$ to an auxiliary Rydberg level $\ket{r_1}$, which can be done by coherent Rabi pulse, creating the superposition of two collective states, each of them having a single Rydberg excitation. Due to Rydberg blockade the second STIRAP pulse will transfer the collective state $\ket{\bar r_0}'$ back to the state $\ket{\bar 0}$ while the state $\ket{\bar r_1}'$ will remain unchanged due to the presence of a single Rydberg excitation in the state $\ket{r_0}$ which blocks the transition $\ket{0} \rightarrow \ket{r_0}$. After the end of the second STIRAP sequence the state $\ket{r_1}$ is transfered to the state $\ket{1}$ by a single $\pi$ pulse.

Pulse areas independent of $N$ on the $\ket{0} \leftrightarrow \ket{r_0}'$ transition can be implemented with STIRAP or ARP
as described above.
We will define the logical basis states as $\ket{\bar 0} = \ket{000...000},$
$\ket{\bar 1} = e^{\imath\chi_N} \ket{\bar 1}',$ and $\ket{\bar r} = e^{\imath\chi_N}\ket{\bar r}'.$ Here $\chi_N$ is
the phase produced by a single $N$-atom STIRAP pulse with positive detuning.  We assume that we do not know the value
of $N$, which may vary from qubit to qubit,  and therefore $\chi_N$ is also unknown, but has a definite value for fixed
$N$.
The logical states are $ \ket{\bar 0}$ and $\ket{\bar 1} = e^{\imath\chi_N} \ket{\bar 1}'$. The auxiliary Rydberg states are defined as 
\bea
\ket{\bar r_0} &=&e^{\imath\chi_N}\ket{\bar r_0}' \\\nonumber
\ket{\bar r_1} &=& e^{\imath\chi_N}\ket{\bar r_1}'.
\eea 
\noindent Starting with a qubit state $\ket{\psi}=a\ket{\bar 0} + b \ket{\bar 1}$
we perform a sequence of pulses 1-5, shown in figure~\ref{Gates}(a), giving 
the sequence of states 
\bea
\ket{\psi_1}&=&a\ket{\bar 0} + ib \ket{\bar r_1}\nonumber\\
\ket{\psi_2}&=&a \ket{\bar r_0} + ib \ket{\bar r_1}\nonumber\\
\ket{\psi_3}&=&a' \ket{\bar r_0} - ib' \ket{\bar r_1}\label{eq.1qubit}\\
\ket{\psi_4}&=&a' \ket{\bar 0} - ib' \ket{\bar r_1}\nonumber\\
\ket{\psi_5}&=&a' \ket{\bar 0} +b' \ket{\bar 1}.\nonumber
\eea

\noindent
The final state $\ket{\psi}=a'\ket{\bar 0} + b' \ket{\bar 1}$ is arbitrary and is selected by the rotation $R(\theta,\phi)$,  in step 3:
$\left({{\begin{array}{*{20}c}
a'\\-b'\\\end{array}}}\right)={\bf R}(\theta,\phi)
\left({{\begin{array}{*{20}c}a\\b\\\end{array}}}\right)$

\textit{CNOT:}
The proposed scheme is an extension of the experiment~\cite{Isenhower2010} and modification of our previous proposal~\cite{Beterov2013}.
Starting with an arbitrary two-qubit state
$\ket{\psi}=a\ket{\bar 0\bar 0} + b \ket{\bar 0 \bar 1} + c\ket{ \bar 1 \bar 0} + d \ket{\bar 1 \bar 1}$
we generate the sequence of states
\bea
\ket{\psi_1}&=&a \ket{{\bar 0}{\bar 0}} + b \ket{{\bar 0} {\bar 1}} +i c\ket{ {\bar r_0}{ \bar 0}} + i d \ket{{\bar r_0} {\bar 1}}\nonumber\\
\ket{\psi_2}&=&a \ket{{\bar 0}{\bar 0}} +i b \ket{{\bar 0} {\bar r_1}} +i c\ket{ {\bar r_0}{ \bar 0}} + i d \ket{{\bar r_0} {\bar 1}}\nonumber\\
\ket{\psi_3}&=&a \ket{{\bar 0}{\bar r_0}} +i b \ket{{\bar 0} {\bar r_1}} + i c\ket{ {\bar r_0}{ \bar 0}} + i d \ket{{\bar r_0} {\bar 1}}\nonumber\\
\ket{\psi_4}&=&i a \ket{{\bar 0}{\bar r_1}}  - b \ket{{\bar 0} {\bar r_0}} - c\ket{ {\bar r_1}{ \bar 0}} - d \ket{{\bar r_1} {\bar 1}}\label{eq.2qubit}\\
\ket{\psi_5}&=&i a \ket{{\bar 0}{\bar r_1}} - b \ket{{\bar 0} {\bar 0}} - c\ket{ {\bar r_1}{ \bar 0}} - d \ket{{\bar r_1} {\bar 1}}
\nonumber\\
\ket{\psi_6}&=& -a \ket{{\bar 0}{\bar 1}} -b \ket{{\bar 0} {\bar 0}} -c\ket{ {\bar r_1}{ \bar 0}} - d \ket{{\bar r_1} {\bar 1}}
\nonumber\\
\ket{\psi_7}&=&-a \ket{{\bar 0}{\bar 1}}  - b \ket{{\bar 0} {\bar 0}} -  i c\ket{ {\bar 1}{ \bar 0}} - i d \ket{{\bar 1} {\bar 1}}
\nonumber.
\eea

\noindent The gate matrix is therefore 
\begin{equation}
U_{CNOT} = \left( {{\begin{array}{*{20}c}
0 & -1 & 0 & 0 \\
-1  & 0 & 0 & 0 \\\label{CNOT}
0  & 0  & -i & 0 \\
0  &  0 & 0 & -i \\
\end{array}} } \right).
\end{equation}
\noindent which can be converted into a standard CNOT gate with a single qubit rotation. 

\textit{Controlled phase gate}
The controlled phase gate is implemented in the way similar to CNOT with replacement of the amplitude-swap sequence by controlled $2\pi$ rotation of the target qubit which 
could be switched on and off by excitation of the control qubit into the Rydberg state. 

We find that arbitrary single qubit rotations in the basis $\ket{\bar 0},\ket{\bar 1}$ can be performed with high
fidelity, without precise knowledge of $N$, by accessing several Rydberg levels $\ket{r_0}$, $\ket{r_1}$ as shown in
figure~\ref{Gates}(a).  Depending on the choice of
implementation, to be discussed below,  this may be given by a one- or two-photon microwave pulse, with Rabi frequency $\Omega_3$.
Provided  states $\ket{r_0}, \ket{r_1}$ are strongly interacting, and
limit the number of excitations in the  ensemble to one, the indicated sequence is obtained.  

The five pulse sequence we describe here is more complicated than the three pulses needed for an arbitrary  single
qubit gate in the approach of Ref.~\cite{Lukin2001}. The reason for this added complexity is that the special phase
preserving property of the double STIRAP or ARP sequences requires that all population is initially in one of the
states connected by the  pulses. The sequence of pulses in figure~\ref{Gates}(a) ensures that this condition is
always satisfied.

All pulses except number 4 in the CNOT
sequence are optical and are localized to either the control or target qubit. Pulse 4 is a microwave field and drives a
$\pi$ rotation on both qubits. As for the single qubit gate the requirement for high fidelity operation is that the
interactions $\ket{r_0} \leftrightarrow \ket{r_0}$, $\ket{r_1} \leftrightarrow \ket{r_1}$, $\ket{r_0} \leftrightarrow
\ket{r_1}$ all lead to full blockade of the ensembles. Since the frequency of pulse 4, which is determined by the energy separation of states
$\ket{r_0},\ket{r_1}$, can be chosen to be very different from the qubit frequency given by the energy separation of
states $\ket{0},\ket{1}$ the application of microwave pulses will not lead to crosstalk in an array of ensemble qubits.

\begin{figure}[!t]
\includegraphics[width=\columnwidth]{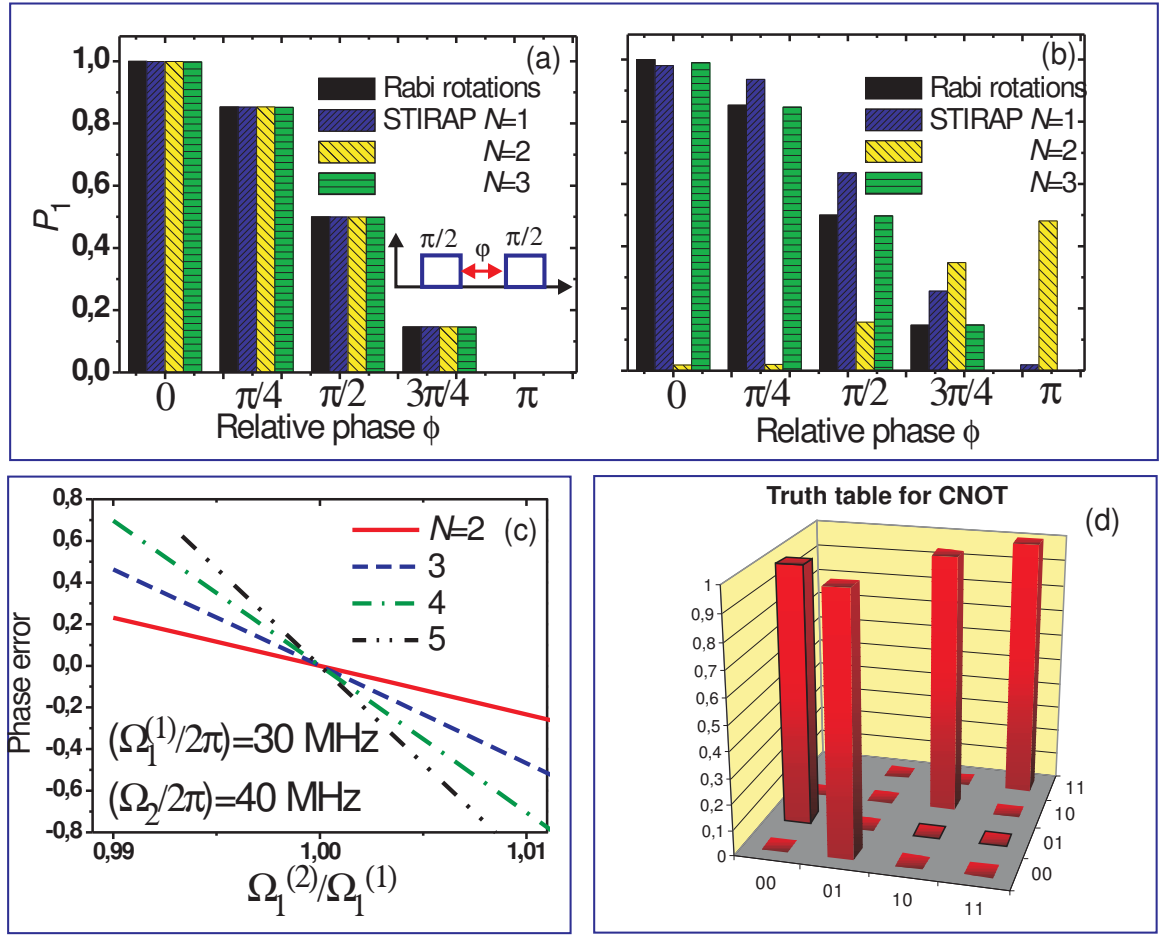}
\vspace{-.5cm}
\caption{
\label{Errors}(Color online).
(a),(b) The dependence of the population of the qubit state $\ket{1}$ after two sequential $\pi/2$ rotations on the phase difference $\phi$ between the pulses with  (a) and without (b) switching the sign of the detuning between the STIRAP sequences; (c) Dependence of the phase error on Rabi frequency changes between pulses for STIRAP; (d) The truth tabel for a CNOT operation described by equation~\ref{CNOT} calculated for a single-atom control qubit and the target qubit which consists of $N=2$ atoms.
}
\end{figure}

\textit{All-optical gates}

Another implementation of single-qubit rotation is based on coherent rotation between the collective states $\ket{\bar r}'$ and $\ket{\bar 1}'$, as shown in figure~\ref{Gates}(d). The subsequent STIRAP sequence will move the state $\ket{\bar r}'$ back to the ground state $\ket{\bar 0}$, while the state $\ket{\bar 1}'$ will be transfered to the state $\ket{\bar r_a}'$ which has an undesirable single-atom Rydberg excitation for $N>1$. This excitation can be removed by an additional $\pi$ pulse which drives the transition $\ket{r} \rightarrow \ket{2}$ where $\ket{2}$ is also a hyperfine magnetic sublevel of the ground state. 

Consider atoms with levels $\ket{0},\ket{1},\ket{2},\ket{r}$ as shown in figure~\ref{Gates}(d). 
We define the ensemble states as:
\bea
\ket{\bar 0} &=&\ket{000...000}\\\nonumber
\ket{\bar 1}' &=& \frac{1}{\sqrt{N (N-1)}}\sum_{j=1}^N\sum_{k \neq j} \ket{000 ... 1_j ...2_k...000}\\\nonumber
\ket{\bar r}' &=& \frac{1}{\sqrt{N}}\sum_{j=1}^N \ket{000 ... r_j ...000}\\\nonumber
\ket{\bar r_a}' &=& \frac{1}{\sqrt{N (N-1)}}\sum_{j=1}^N\sum_{k \neq j} \ket{000 ... 1_j ...r_k...000}.
\eea
\noindent
Similarly to the previous discussion, we define the logical basis states as $\ket{\bar 0} = \ket{000...000},$
$\ket{\bar 1} = e^{\imath\theta_N} \ket{\bar 1}'$, $\ket{\bar r_0} = e^{\imath\chi_N}\ket{\bar r_0}'$ and $\ket{\bar r_a} = e^{\imath\theta_N}\ket{\bar r_a}'.$ 
Here we take into account that the dynamic phase $\theta_N$, arised after two STIRAP sequences, is different from the phase $\chi_N$.

The logical states are $ \ket{\bar 0}$ and $\ket{\bar 1} = e^{\imath\theta_N} \ket{\bar 1}'$. 
The requirement to use three states $\ket{0}, \ket{1}$ and $\ket{2}$ with low decoherence for logical encoding could be a limiting factor for experimental implementation of this all-optical scheme compared to the microwave gates discussed above, where only two hyperfines sublevels were necessary. The advantage of this scheme is the elimination of the need to use microwave radiation to drive coherent transitions between the Rydberg states which could be challenging in experiment.

The auxiliary Rydberg states are defined as 
\bea
\ket{\bar r} &=&e^{\imath\chi_N}\ket{\bar r}' \\\nonumber
\ket{\bar r_a} &=& e^{\imath\theta_N}\ket{\bar r_a}'.
\eea

\noindent Starting with a qubit state $\ket{\psi}=a\ket{\bar 0} + b \ket{\bar 1}$
we perform a sequence of pulses 1-5, shown in figure~\ref{Gates}(d), giving 
the sequence of states 
\bea
\ket{\psi_1}&=&a\ket{\bar 0} + ib \ket{\bar r_a}\nonumber\\
\ket{\psi_2}&=&a \ket{\bar r_0} + ib \ket{\bar r_a}\nonumber\\
\ket{\psi_3}&=&a' \ket{\bar r_0} - ib' \ket{\bar r_a}\label{eq10}\\
\ket{\psi_4}&=&a' \ket{\bar 0} - ib' \ket{\bar r_a}\nonumber\\
\ket{\psi_5}&=&a' \ket{\bar 0} +b' \ket{\bar 1}.\nonumber
\eea
\noindent
The CNOT gate shown in figure~\ref{Gates}(d) and controlled phase gate shown in figure~\ref{Gates}(e) are equivalent to the corresponding microwave gates.
To verify that our scheme preserves coherence, we have numerically modelled the sequence of two single-qubit rotations
for an angle of $\pi/2$ with relative phases $\phi$ in the range $0-\pi$. The probability to find the ensemble in the
qubit state $\ket{1}$  was calculated for our STIRAP-based protocol for $N=1-3$ atoms  and compared with the outcome of
similar single-atom gate sequence applied using conventional Rabi rotations [shown as black in figure~\ref{Errors}(a)]. We have
found that the probability for the ensemble to be in state $\ket{1}$ is independent of the number of atoms and
correctly depends on the relative phase between the microwave pulses, as shown in Fig.~7(a). In contrast, if we do not
switch the detuning from the intermediate state after the first STIRAP pulse, the probability to find the ensemble in
the state $\ket{1}$  becomes $N$-dependent and is inconsistent with the expected values, as shown in figure~\ref{Errors}(b).

Although the proposed double-pulse sequences are almost insensitive
 to moderate variations of the absolute Rabi frequency, the main sources of errors are fluctuations of the Rabi frequencies between
 the first and second pulses. For perfectly identical pulses the population transfer error in ensembles of $N=5$ atoms can be kept below $10^{-3}$ for STIRAP and below $10^{-4}$ for
 an ARP pulse  for a wide range of Rabi frequencies.
The dependence  of the phase errors on parameters of the laser pulses are shown in figure~\ref{Errors}(c). The
dependence of the phase error on the ratio of Rabi frequencies $\Omega_1^{(2)}/\Omega_1^{(1)}$ between pulses [see
figure~\ref{DoubleSequence}(b)] is shown in figure~\ref{Errors}(c) for $N=1-5$ atoms.  The single-photon ARP excitation  in
demonstrates reduced sensitivity to  fluctuations of the Rabi frequency and has higher efficiency
at lower Rabi frequencies. Although this could be an important advantage over STIRAP, implementation of single-photon
Rydberg excitation is difficult due to the need of ultraviolet laser radiation and larger sensitivity to Doppler
broadening~\cite{Ryabtsev2011,Entin2013}. For either approach the double pulse amplitudes  must be well matched for low phase
errors. Using a fiber delay line amplitude matching at the level of $10^{-6}$ is feasible over the timescale of few
microseconds~\cite{Wineland1998}.

Figure~\ref{Errors}(d) presents the numerically calculated truth table for a microwave CNOT operation for a single control atom and a target ensemble which consists of one or two atoms. The accuracy better than $0.3\%$ has been obtained from the Shr\"odinger equation without taking into account decay of the intermediate and Rydberg states. This proves that high efficiency of two-qubit logic gates could be achieved using the schemes which we have proposed. The decoherence caused by the finite lifetime of the Rydberg and intermediate excited states~\cite{Beterov2009} could affect the fidelity of quantum gates, but our previous simulations using density matrix approach have confirmed that for realistic experimental parameters the errors could be small enough~\cite{Beterov2013}.

\section{Conclusion}
We have developed the schemes for coherent control of mesoscopic atomic ensembles based on single-photon and two-photon adiabatic passage and Rydberg blockade.
These schemes allow for deterministic single-atom loading of optical dipole traps and single-qubit and two-qubit quantum logic gates.

\section*{Acknowledgements}
This work was supported by the grant of the President of Russian Federation MK.7060.2012.2, EPSRC project EP/K022938/1, RAS, RFBR Grants No. 13-02-00283 and 14-02-00680, the project
FP7-PEOPLE-2009-IRSES-COLIMA and Russian Quantum Center.  MS was supported by the NSF and  the AFOSR MURI program.

\section*{References}

\providecommand{\newblock}{}

\end{document}